\documentclass[fleqn,usenatbib]{mnras}
\usepackage{graphicx,epsfig,psfig,multicol}
\usepackage[dvipsnames]{xcolor}
\usepackage[]{inputenc,amssymb}
\usepackage{amsmath}
\usepackage{color}
\usepackage{textcomp}
\usepackage{xcolor,cancel}
\usepackage{comment,float}
\usepackage[export]{adjustbox}

\usepackage{float}

\usepackage{lipsum}
\usepackage{mwe}
\usepackage{arydshln} 

\definecolor{webgreen}{rgb}{0,.5,0}
\definecolor{webbrown}{rgb}{.6,0,0}

\usepackage[toc,page]{appendix}

\def \beq{\begin{equation}}
\def \eeq{\end{equation}}
\def \bea{\begin{eqnarray}}
\def \eea{\end{eqnarray}}

\usepackage[T1]{fontenc}

\DeclareRobustCommand{\VAN}[3]{#2}
\let\VANthebibliography\thebibliography
\def\thebibliography{\DeclareRobustCommand{\VAN}[3]{##3}\VANthebibliography}

\title[DM Heating of Clusters]{Heating Galaxy Clusters with Interacting Dark Matter}

\author[Y. Shoji et al.]{
Yutaro Shoji\thanks{E-mail: yutaro.shoji@mail.huji.ac.il},
Eric Kuflik,
Yuval Birnboim,
Nicholas C. Stone
\\
Racah Institute of Physics, The Hebrew University, 91904, Jerusalem, Israel}

\date{Accepted XXX. Received YYY; in original form ZZZ}

\pubyear{2023}

\begin{document}

\label{firstpage}
\pagerange{\pageref{firstpage}--\pageref{lastpage}}
\maketitle

\begin{abstract}

The overcooling of cool core clusters is a persistent puzzle in the astrophysics of galaxy clusters.  We propose that it may naturally be resolved via interactions between the baryons of the intracluster medium (ICM) and its dark matter (DM).  DM-baryon interactions can inject heat into the ICM to offset bremmstrahlung cooling, but these interactions are also strongly constrained by existing experiments and astrophysical observations.  We survey existing constraints and combine these with the energetic needs of an observed sample of cool core clusters.  We find that a robust parameter space exists for baryon-DM scattering solutions to the cooling flow problem, provided that only a sub-component of DM interacts strongly with the baryons.  Interestingly, baryon-DM scattering is a thermally stable heating source so long as the baryon temperature is greater than $1/3-1/2$ the DM temperature, a condition that seems to be satisfied observationally.

\end{abstract}

\begin{keywords}
galaxies: clusters: general -- galaxies: clusters: intracluster medium -- (cosmology:) dark matter
\end{keywords}

\twocolumn

\section{Introduction}

Galaxy clusters are the largest gravitationally bound systems in the Universe.  Their unique size makes them valuable laboratories for understanding the Universe's cosmological expansion history \citep{Sunyaev:1972, Ostriker:1986, Vikhlinin:2009a, Vikhlinin:2009b}, for testing theories of dark matter \citep{Markevitch:2004, Clowe:2006}, and for probing our understanding of relativistic gravity \citep{Wojtak:2011}.  The deep potential well of cluster environments also has many astrophysical consequences, such as enhanced rates of galaxy interactions \citep{Moore:1996, Moore:1998}, ram pressure stripping \citep{Gunn:1972}, and the formation of the largest supermassive black holes in the Universe \citep{Mittal:2009}. 

One persistent astrophysical puzzle concerns the thermodynamic evolution of galaxy clusters.  The space between galaxies in a cluster, which includes about $90\%$ of the cluster baryons, is inhabited by the intracluster medium (ICM), a low-density, high-temperature plasma that steadily releases energy through free-free (bremsstrahlung) radiation.  In an order unity fraction of galaxy clusters, the central densities and temperatures of the ICM are high enough that the plasma should be able to cool appreciably in much less than a Hubble time. About half of observed clusters, the ``cool core'' (CC) clusters, have central cooling times of about $300$~Myr -- much shorter than the Hubble time \citep{cavagnolo09}. These also have central ICM temperatures that are a factor $\approx 2-3$ lower than in outer ICM regions.  Since the cooling time for the ICM decreases with decreasing temperature, this process should run away and lead to massive inflows of cold gas (``cooling flows;'' \citealt{Fabian:1977, Cowie:1977}) that impact the central brightest cluster galaxy (BCG) and trigger high rates of star formation ($\sim 10^3 M_\odot~{\rm yr}^{-1}$; \citealt{Fabian:1994}).  In reality, the actual amounts of observed central cold gas are far below the cooling flow predictions \citep{Peterson:2003}, and likewise the level of observed central star formation is only $\sim 1-10\%$ the level that is predicted from simple cooling flow models \citep{Fabian:1994, McDonald:2018}.  These discrepancies between theory and observation comprise the ``cooling flow problem'' (see \citealt{McNamara:2012, Donaghue:2022} for relevant reviews).

Various astrophysical solutions to the cooling flow problem have been considered, many of which run into serious obstacles.  For example, thermal conduction via electrons was once seen as a promising way to connect the rapidly cooling cluster core with the larger reservoir of thermal energy contained in the baryonic outskirts of the cluster \citep{Voigt:2002, Zakamska:2003}.  However, in the magnetized environment of the ICM, conduction is strongly suppressed in directions perpendicular to magnetic field lines, and convective instabilities present in the ICM are likely to reorient magnetic field lines into a tangential direction, strongly suppressing radial conduction \citep{Parrish:2009, Parrish:2012}.  Another possibility is gravitational heating due to the backreaction from drag forces that decelerate massive streams or clumps infalling into the center of the ICM \citep{Dekel:2008, Birnboim:2011, Zinger:2018}.  Today, the leading solution to the cooling flow problem is feedback from active galactic nuclei (AGNs) in the galaxy cluster \citep{Binney:1995, Quilis:2001, DallaVecchia:2004, Sijacki:2006}.  The inflow of even small amounts of gas into the BCG or other centrally located galaxies will initiate accretion onto the supermassive black holes (SMBH) of these galaxies.  The high efficiency with which accretion converts rest mass into bulk kinetic energy (via production of relativistic jets and sub-relativistic winds) allows the accreting AGN to inject large amounts of heat into bubbles surrounding the accreting galaxy.  Indeed, high-resolution {\it Chandra} and {\it XMM-Newton} observations find evidence for these bubbles around BCGs in many CC galaxies \citep{dunnRadioBubblesClusters2005,zhangGenerationInternalWaves2018}.  

However, the AGN solution to the cooling flow problem is not without its own difficulties.  For AGN feedback to compensate for the cooling in CC clusters, the SMBH must inject energy very smoothly, both temporally and spatially. If the SMBH shuts off for a cooling time, runaway cooling would resume, leading to complete cooling of the central ICM, and possibly to an energetic burst of accretion feedback from the SMBH that heats and expels the ICM. Spatial uniformity is also required: for example, if the heating is mainly done by AGN jets, material perpendicular to the jets could still cool and make stars. Energy injection must be smooth on scales
larger than the distances over which 
heat transport is slower than cooling.
Since conduction is likely inefficient at transporting heat \citep{parrishMagnetothermalInstabilityIntracluster2008,parrishAnisotropicThermalConduction2009}, turbulent diffusion of hot material into cooling material would be the main heat transport mechanism. Anisotropy of emitted X-rays as well as line broadening both indicate that turbulent velocities in the ICM are a few hundreds of ${\rm km~s^{-1}}$ \citep[see, for example,][]{churazov12}, demonstrating that turbulent diffusion can only be significant for ${\rm 20~kpc}$ distances or less. Indeed, hydrodynamic simulations attempting to recover the CC cluster fraction and properties, have had only moderate success in reproducing the observations, and are sensitive to the AGN feedback process \citep[see, for example,][using state-of-the-art IllustrisTNG simulations]{barnes18}. The turbulent mixing depends strongly on the driving mechanism of the turbulence, which is sensitive to the physical nature of the AGN jet \citep{hillelGentleHeatingMixing2017,yangHOWAGNJETS2016}. A systematic analysis of the necessary properties of AGN feedback to be a successful solution to the overcooling problem is presented in \citet{gaspariSolvingCoolingFlow2013,gaspariLinkingMacroMeso2020}. While these feedback models show promise, a considerable amount of fine-tuning of the feedback is still necessary. Purely radiative AGN heating is expected to be thermally unstable; mechanical heating processes are more complex and the thermal stability of their ICM equilibria remains unclear \citep{mccourtThermalInstabilityGravitationally2012,sharmaThermalInstabilityFeedback2012}. Variants of AGN feedback models pump energy into the IGM by shocks or Landau damping of sound waves \citet{fabianVeryDeepChandra2006}, or by cosmic rays \citep{guoFeedbackHeatingCosmic2008a}.

Aside from these astrophysical solutions to the cooling flow problem, a handful of alternative scenarios have been proposed invoking new physics, primarily novel interactions between baryons and particle dark matter (DM).  \citet{Qin:2001} suggested that a sufficiently large cross-section for baryon-DM interactions could allow ICM baryons to access the thermal reservoir of cluster DM, providing a heat source that can offset free-free cooling.  Although this original proposal was later shown to be thermally unstable, \citet{Chuzhoy:2006} showed how some energy-dependent cross-sections could allow baryon-DM heating to achieve a stable thermal balance. 

Since these early works on cluster heating, many more constraints have emerged that limit possible cross-sections for baryon-DM interactions, in a broader range of phenomenological models for particle DM.
In particular, an efficient baryon-DM scattering before recombination is strongly constrained by the Cosmic Microwave Background (CMB), the Baryon Acoustic Oscillations (BAO), and other matter power spectrum observations. These constraints are strong enough to exclude all the relevant parameter space if {\it all} of the DM particles 
interact strongly with the baryons.  To avoid these constraints, we consider multi-component DM and assume that only a small portion of DM has a large scattering cross-section with baryons, which results in very different phenomenological consequences in comparison to the original work.
Another development that limits possible interaction cross-sections is the observation of cooling rates in other astronomical objects. Since baryon-DM interactions should also heat up such objects, these observations give more direct constraints on how much heating can occur~\citep{Wadekar:2019mpc}.

In this paper, we revisit the hypothesis that baryon-DM interactions may resolve the cooling flow problem in light of these recent developments.  In \S \ref{sec:heating}, we
describe the observational datasets we use to test our DM heating model.
In \S \ref{sec:scatter}, we develop a simple but general model for particle DM that heats ICM baryons through two-body scattering, and we show that in contrast to previous work, this heating mechanism can be thermally stable over astrophysically relevant temperature ranges.  (In Appendix \ref{sec:dp}, we perform the same analysis for another model with dark photon DM, which turns out to be excluded.)  In examining the viability of the DM heating model, we thoroughly explore the impact of laboratory and astrophysical constraints (with a fuller discussion in Appendix \ref{sec:constraints}), and we summarize in \S \ref{sec:conclusions}.  Throughout the paper, we use a units system where the reduced Planck constant, speed of light, and Boltzmann constant are unity ($\hbar = c = k_{\rm B}=1$).

\section{Galaxy cluster profiles}
\label{sec:heating}
Estimating the heating due to DM-baryonic interactions requires knowledge of the DM density profile, the velocity dispersion of the DM (its ``temperature''), and the number density of the DM, as well as the density and temperature profiles of the baryons in the ICM. The net radiative cooling of the baryons requires knowledge of the baryonic density and temperature \citep[as well as its metallicity, though this contributes less at higher temperature and is only a mild correction above $1$ keV;][]{sutherlandCoolingFunctionsLowDensity1993,gnatTimeDependentIonizationRadiatively2007}.

In the absence of direct measurements of the DM component, we estimate its properties from universal profiles for DM halos that are based on their mass, redshift and formation histories \citep{navarroUniversalDensityProfile1997}. We note that while these universal NFW profiles have been very successful in reproducing observations and numerical simulations, a large amount of cosmic variance exists, introducing significant scatter. 

The baryonic density and temperature profiles of clusters can be observed by their X-ray emission and by the Sunyaev-Zel'dovich effect \citep{sunyaevSmallscaleFluctuationsRelic1970a}. However, these measurements become exceedingly difficult at the very centers of clusters, where the ICM becomes sensitive to the BCG and to AGN activity that pushes it away from radial symmetry and temporal stability. In this work, we require robust, reasonably accurate cooling rates of the baryons near the centers of CC clusters. \citet{cavagnolo09} compiled entropy profiles and cooling times of a large sample of {\it Chandra} and HIFLUGCS (XMM and ROSAT) observed clusters. They find that entropy cores are ubiquitous in clusters below $100$~kpc, and that the central entropy value is bi-modal, separating the CC clusters ($K_0\le 30-50~\rm{keV\,cm^2}$) and the non-cool core (NCC; $K_0>50~\rm{keV\,cm^2}$). However, their data does not include good total mass estimates, as is needed to deduce the DM profiles for our analysis. To this end, we use the mass estimates from  
\citet{Andrade-Santos_2021} that compares mass estimates of clusters, based on observations from PLANCK Early Sunyaev-Zel'dovich \citep{collaborationPlanck2015Results2016} and from archival {\it Chandra} X-ray observations. We assume NFW profiles that recover their observed $M_{500}$ (the mass encompassed in a sphere with an average density $500$ times the universal density) with concentration parameters of $c_{\text{NFW}}=4$, as is appropriate for low-redshift clusters \citep{wechslerConcentrationsDarkHalos2002}. 

Combining the entropy profiles and the mass estimates, and limiting ourselves to CC clusters (based on the central entropy criterion), we are left with 14 clusters: ABELL 0085, ABELL 0478, ABELL 0496, ABELL 1644, ABELL 1795, ABELL 2029, ABELL 2199, ABELL 2204, ABELL 2390, ABELL 3112, ABELL 3528S, 2A 0335+096, RX J1720.1+2638, and ZWCL 1742.
In the rest of this work, we will study the nature and amplitude of the DM-baryon interactions that will be required to offset the cooling of these 14 clusters. 

\section{DM-baryon scattering}
\label{sec:scatter}
We consider clusters that are heated by energy transfer from a DM relic, $\chi$, to baryons, $B$, through pairwise collisions.
For the energy transfer rate between two fluids, we follow the calculations in \cite{Dvorkin:2013cea,Gluscevic:2017ywp,Xu:2018efh,Slatyer:2018aqg,Boddy:2018wzy}. When the relative bulk velocity between the baryon and $\chi$ fluids is negligible, the energy transfer rate from $\chi$ to baryons per unit volume is given by
\begin{equation}
    \frac{dQ_{\chi\to b}}{dt}=3(T_\chi-T_b)\sum_B\frac{\rho_\chi\rho_B}{(m_\chi+m_B)^2}\sigma_n^{\chi B}c_nu_B^{n+1},\label{eq_energy_transfer}
\end{equation}
where $\rho_i$, $T_i$ and $m_i$ are the mass density, the temperature and the mass of component $i$, respectively, $c_n=\frac{2^{(5+n)/2}}{3\sqrt{\pi}}\Gamma(3+n/2)$, $u_B^2=T_\chi/m_\chi+T_b/m_B$ and $B$ runs over the baryon species.
The velocity-stripped momentum-transfer cross section of $\chi -B$ scattering, $\sigma_{n}^{\chi B}$, is
\begin{equation}
    \sigma_{n}^{\chi B}v_{\rm rel}^n \equiv \int d\cos\theta_*\frac{d\sigma^{\chi B}}{d\cos\theta_*}(1-\cos\theta_*),
\end{equation}
where $v_{\rm rel}$ is the relative velocity and $\theta_*$ is the scattering angle in the center-of-mass frame. Note that the cross section with the lowest $n$ dominates the cross section for low-velocity scattering.
We take into account H and He, with $n_{\rm He}/n_{\rm H}=0.08$. We will focus on $n=0$ as a fiducial model, corresponding to contact interactions (heavy mediator). Coloumb-like interactions give $n=-4$ (massless mediator), while interactions akin to an electric dipole moment give $n=-2$.

We assume the velocities of $\chi$ and $\psi$ obey the thermal distribution for simplicity. In reality, the velocity distribution of $\psi$ may be distorted around the center of the galaxy cluster. This however does not impact our results much because, with the same cross section of $\chi$-$\psi$ interaction as that of the $\chi$-baryon interaction, the free streaming length of $\chi$ is comparable to the size of galaxy clusters. Thus, $\chi$ is not sensitive to the local distortion of the $\psi$ velocity distribution. Meanwhile, the velocity distribution of $\chi$ can be distorted by the continuous heating and cooling by $\psi$ and baryons. In this paper, we approximate it with a thermal distribution with a single temperature and leave a more precise treatment as a future work.

On the other hand, the cooling rate of a cluster is characterized by the cooling time,
\begin{equation}
    t_c^{-1}\equiv \Big|\frac{d\ln \varepsilon}{dt}\Big|= -\frac{d\ln T_b}{dt}.
\end{equation}
where $\varepsilon$ is the internal energy per unit volume. It can be translated into the energy loss per unit volume as
\begin{equation}
    \label{eq:dQraddT}
    \frac{dQ_{\rm rad}}{dt}=-\frac{3}{2}n_b\frac{T_b}{t_c},
\end{equation}
where $n_b$ is the baryon number density.
To maintain a thermal equilibrium (i.e. a static balance between heating and cooling rates) in the cluster, we require
\begin{equation}
    \frac{dQ_{\rm rad}}{dt}+\frac{dQ_{\chi\to b}}{dt}\simeq0,
\end{equation}
in the cores of the clusters.

Next, we estimate $T_\chi$.
We do not assume that $\chi$ explains the whole DM energy density $\Omega_{\rm DM}$, but only a fraction thereof. We define the ratio as
\begin{equation}
    f_\chi=\frac{\Omega_\chi}{\Omega_{\rm DM}},
\end{equation}
and for simplicity assume a uniform fraction $f_\chi$ everywhere in all galaxy clusters.
Again for simplicity, we consider two-component DM where the DM energy density is dominated by a different dark particle, $\psi$, and assume that $\chi$ and $\psi$ are in a  kinetic equilibrium. Then, from the virial theorem,
\begin{equation}
    T_\chi=T_\psi\simeq0.3\frac{GM}{R}\left(\frac{f_\chi}{m_\chi}+\frac{1-f_\chi}{m_\psi}\right)^{-1}, \label{eq:equipartition}
\end{equation}
where $G$ is the gravitational constant, $M$ is the virial mass and $R$ is the virial radius of the cluster. The prefactor to the binding energy, set by the internal structure of the cluster's profile, is approximated from \citet[][sec. 9.3]{binneyGalacticDynamics1987}. We note that the specific prefactor will scale the DM temperature and the baryonic temperature in a similar manner and will not affect the validity of the analysis.

Here, $T_\chi$ needs to be higher than $T_b$ at the core region of CC clusters to heat up the baryons. Since $T_b$ at the core region is typically lower than the virial temperature, the minimum value of $m_\psi$ is slightly smaller than the proton mass. 
Regardless of how large the the interaction cross-section $\sigma_n^{\chi B}$ becomes, this heating mechanism will not over-heat baryons in NCC clusters because in this case the baryons will reach thermal equilibrium with the $\chi$ (and $\psi$) particles.  This thermodynamic conclusion can also be seen explicitly from the functional form of Eq. \ref{eq_energy_transfer}.

Various observations constrain the same baryon-DM cross-section that may explain the cool core puzzle. Since the cross-section may dependent on the momentum transfer $q$, the cross-section for the same fundamental interaction depends on the measurement being performed. Therefore we define the cross-section
\begin{equation}
\bar{\sigma}_n^{\chi B} \equiv \frac{\overline{|\mathcal{M}|^2}}{16\pi (m_B +m_\chi)^2}\bigg|_{q=q_{\rm ref}}
\end{equation}
evaluated at a fixed momentum ${q_{\rm ref}= \alpha m_e}$. Here $\overline{|\mathcal{M}|^2}$ is the matrix element squared and summed/averaged over internal degrees-of-freedom. In terms of the velocity-stripped momentum-transfer cross section 

\begin{align}
    \sigma_{-4}^{\chi B}&\simeq\frac{\bar\sigma_{-4}^{\chi B}}{8}\frac{q_{\rm ref}^4}{\mu_{\chi B}^4}\ln\frac{4 \mu_{\chi B}^2v_{\rm rel}^2}{m_D^2},\label{sigma-4}\\
    \sigma_{-2}^{\chi B}&=\frac{\bar\sigma_{-2}^{\chi B}}{2}\frac{q_{\rm ref}^2}{\mu_{\chi B}^2},\\
    \sigma_0^{\chi B}&=\bar\sigma_{0}^{\chi B},
\end{align}
where 
$\mu_{\chi B} = \frac{m_\chi m_B}{m_\chi +m_B}$ is the DM–baryon reduced
mass and $m_D=8\pi\alpha n_b/T_b$ is the Debye screening mass \citep{Buen-Abad:2021mvc}.

A detailed discussion of the relevant observational and experimental constraints is given in the Appendix \ref{sec:constraints}, but we summarize them here. Strong baryon-DM coupling can have observable effects on the CMB and matter power spectrum. These constraints are very strong, and exclude all relevant parameter space if $\chi$ makes up all of the DM ($f_\chi =100\%$). On the other hand, the constraints disappear for $f_\chi < 0.4\%$, corresponding to roughly the error in the baryonic density measurement. However, even with $f_\chi \ll 1$, avoiding this type of CMB constraint generally requires one extra assumption on either baryon-$\chi$ or $\psi$-$\chi$ interactions; we survey three possibilities in the Appendix \ref{apx_avoid_cmb}.

\begin{figure*}
    \centering
    \includegraphics[width=0.48\linewidth]{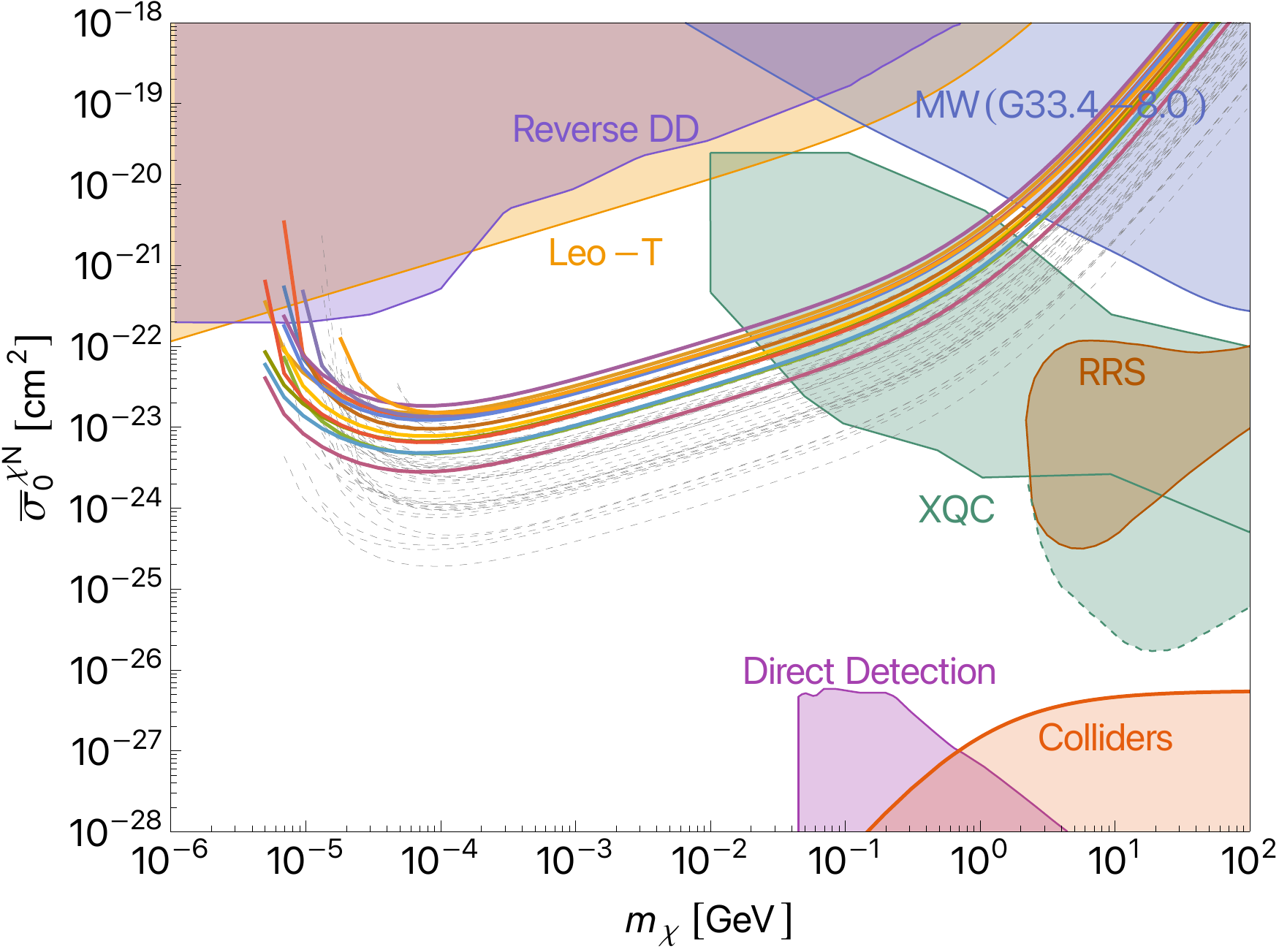}     \hfill \includegraphics[width=0.48\linewidth]{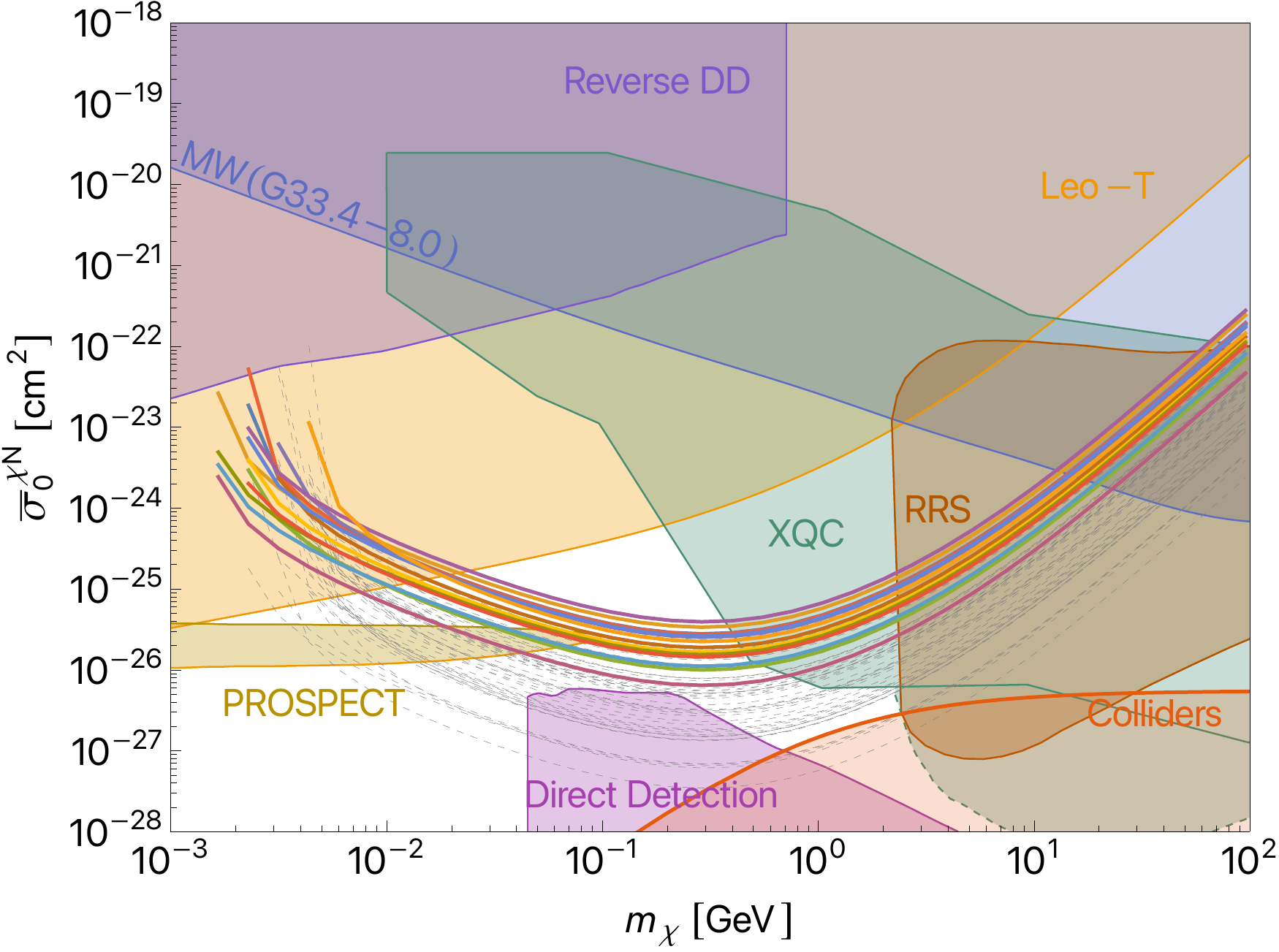}
    \caption{Allowed parameter region and the cross section required to explain CC clusters with baryon-DM scattering. Nucleon-only coupling, {\bf left:} $m_\psi=3~{\rm GeV}$, $f_\chi=0.001\%$; {\bf right:} $m_\psi=100~{\rm GeV}$, $f_\chi=0.4\%$.
    The solid lines show the required cross section for DM-baryon scattering to heat the 14 CC clusters at the same rate as they lose energy from bremsstrahlung cooling. The dashed lines show the same condition on NCC clusters. Also shown are the relevant laboratory and astrophysical observations constraints on the parameter space; see Appendix \ref{sec:constraints} for more details. 
 All curves and constraints assume a velocity-independent cross-section ($n=0$).  Substantial parameter space exists for baryon-DM scattering to offset bremsstrahlung cooling in realistic ICM. 
    }
    \label{fig:nucleus}
\end{figure*}

DM direct-detection experiments also constrain the same interaction. These  can provide very strong limits on the coupling strength. However, if the DM interacts too strongly with the atmosphere or Earth's crust, then the DM may never make it to the detector, and is unconstrained. Therefore, surface runs of CREST-III \citep{CRESST:2017ues} and EDELWIESS \citep{EDELWEISS:2019vjv} provide the strongest constraints.  Rocket and balloon experiments, such as the X-ray Quantum Calorimetry (XQC) experiment~\citep{McCammon:2002gb} provide a possible probe for very strong interactions, but they are also not completely free of shielding.

An additional consequence of strong interactions with the baryons is scattering with cosmic rays. First, the DM can be accelerated by scattering off high energy cosmic rays in the Milky Way. The boosted DM can then be detected by DM direct detection, but also by
neutrino experiments (which are typically not sensitive enough to find slow-moving, virialized DM: \citealt{Yin:2018yjn,Bringmann:2018cvk, Ema:2018bih}). Conversely, DM scattering with cosmic rays can alter the cosmic ray spectrum in a measurable way, leading to constraints known as `reverse direct detection'~\citep{Cappiello:2018hsu}.

The same scattering interactions that we invoke here to heat ICM baryons can also lead to DM pair-production in colliders, and this can be searched for with missing transverse energy signals. For a very strong interaction, however, the dark particles interact with the calorimeter and the constraint does not apply~\citep{Daci:2015hca}.

The heating mechanism for the CC clusters also heats up other astronomical objects and can be probed as discussed in  \citet{Wadekar:2019mpc}. The most relevant constraints come from the gas-rich dwarf Leo-T and Milky Way gas clouds (G33.4-8.0). The constraints presented by \citet{Wadekar:2019mpc} are based on the observation of cool gas ($6000~{\rm K}$ for Leo-T and $400~{\rm K}$ for G33.4-8.0), that would have been heated if DM interactions were strong enough. However, the analysis there depends on some assumptions: that the baryons (gas) is relaxed and is not in the process of being heated, that the gas is in hydrostatic pressure equilibrium, that the gas density is smooth rather than clumped (since the cooling rate depends on the density squared), and that the dark matter profiles are spherical and follow a Burkert profile. While these assumptions are reasonable, considerably larger statistics, and more careful analysis are necessary before the constraints can be treated as strong. Regardless, we plot them here to demonstrate the potency of such arguments to limit the DM cross-sections, when better statistics or better understanding of these systems is available.

We also note that in this analysis we are ignoring additional heating terms that originate from the interaction between the Galactic CGM and gas clouds, and Leo-T and the local group's halos. These terms can originate from  photoionizing radiation   \citep[both  the background and because of the vicinity to the Galaxy][]{haardtRadiativeTransferClumpy2012} or from shocks. Ignoring these terms is an upper-bound for the amount of heating needed by the DM interactions. Also, cooling can also be enhanced by the formation of mixing layers between the clouds and the CGM, and non-equilibrium cooling \citep{gnatTimeDependentIonizationRadiatively2007}. These processes are very complex \citep{fieldingMultiphaseGasFractal2020}, are very sensitive to the nature of feedback from the Galactic disc, and are left for future work using high-resolution numerical simulations.

In Fig.~\ref{fig:nucleus}, we show the $\{m_\chi, \bar{\sigma}_0^{\chi N}\}$ parameter space that can resolve the cool core puzzle, along with the observational and experimental constraints discussed above.  We plot this parameter space for the $n=0$ case alone, because (i) combinations of $m_\chi$ and $\bar{\sigma}_0^{\chi N}$ that can resolve the cool core problem are ruled out by Leo-T observations \citep{Wadekar:2019mpc} for $n=-2$ and $n=-4$, and (ii) many of the other observational constraints have been computed for $n=0$ scatterings and can not be translated trivially to generic $n$.  In Fig. \ref{fig:nucleus}, we see that a broad parameter space exists for $n=0$ $\chi$-baryon interactions to heat CC clusters.  For example, if $m_\psi=3$ GeV and $f_\chi=10^{-5}$, a broad range of $\chi$ masses ($10^{-5} \lesssim m_\chi / {\rm GeV} \lesssim 10^{-2}$) and scattering cross-sections ($10^{-23} \lesssim \bar{\sigma}_0^{\chi N} / {\rm cm}^2 \lesssim 10^{-22}$) are able to supply sufficient heat to prevent runaway ICM cooling in CC clusters.  Large $\chi$ masses are ruled out by experimental constraints (primarily XQC), while small $\chi$ masses eventually cause $T_\chi$ to fall below the baryon temperature (see Eq. \ref{eq:equipartition}), rendering baryon-DM interactions a heat sink rather than a heating source.  Other combinations of $m_\psi$ and $f_\chi$ will increase or reduce the available parameter space to resolve the cool core problem; a less favorable example ($m_\psi = 100$ GeV and $f_\chi  = 4\times10^{-3}$) is also shown in Fig. \ref{fig:nucleus}.

Additional model-dependent constraints (and internal self-consistency issues) exist, arising due to large DM self-interactions that thermalize the two components of DM in the halo.  Such observations do not constrain the DM-baryon scattering rate, and these need to be treated in a given model of DM. We discuss a possible realization of the DM self-interactions and their consequences in Appendix~\ref{ref:DMself}.

The two-component DM model introduced in this paper has an effective self-interaction that may have observable implications for galaxy-scale dynamics, akin to the ``partially interacting'' DM of \cite{2013PDU.....2..139F}.  Self-interacting DM models are known to have a wide variety of dynamical signatures: for example, dissipative self-interactions may lead to the formation of condensed, low-aspect ratio DM disks (e.g. \cite{2013PDU.....2..139F,2013PhRvL.110u1302F}), although this is not necessarily implied in our scenario due to the conservative interactions we invoke for the $\chi$ particle.  However, even though conservative self-interactions are unlikely to form dynamically cold disks of DM, they can still lead to collisional evolution of DM halos.  If the relaxation time of a DM halo due to self-interactions is shorter than a Hubble time, then a collisionless equilibrium set by violent relaxation (e.g. a Navarro-Frenk-White profile,  \cite{1997ApJ...490..493N}) can evolve over time into a quasi-isothermal collisional equilibrium, in analogy to the core collapse phenomenon in globular clusters (\cite{2002PhRvL..88j1301B,2002ApJ...568..475B,2015MNRAS.453...29E,2023MNRAS.523.4786O}).  While such core-collapsed DM halos could in principle be distinguished observationally from standard ones, we defer a detailed study of this possible constraint for future work.

\subsection{Stability}
\label{sec_stability}

A thermodynamic equilibrium where the volumetric heating rate (in this case $q_\chi = {\rm d}Q_\chi/{\rm d}t$) balances cooling ($q_{\rm rad}={\rm d}|Q_{\rm rad}|/{\rm d}t$)  will only be stable if $\partial q_\chi / \partial T_b < \partial q_{\rm rad}/\partial T_b$.  Both of these partial derivatives are evaluated assuming an isobaric medium, i.e. one where the local sound crossing time is much shorter than the cooling time.  Assuming an ideal gas equation of state, the isobaric assumption implies that the baryon density $\rho_{\rm B} = \mu_{\rm M} m_{\rm p} P / (kT_b)$, with $\mu_{\rm M}$ the mean molecular weight of the baryons and pressure $P$ constant.

Taking our fiducial case of $n=0$, we collect constants and rewrite $q_{\rm rad}=A_{\rm rad} \rho_{\rm B}^2 T_b^{1/2}$ and $q_{\chi} = A_\chi (T_\chi - T_b) \rho_{\rm B} \sqrt{T_\chi / m_\chi + T_b/m_{\rm B}}$ (for simplicity, we have assumed the existence of only a single baryon species in this calculation).  Both $A_{\rm rad}$ and $A_{\chi}$ are constants that do not depend on the hydrodynamic variables $\rho_{\rm B}$ and $T_b$; likewise, we assume that $T_\chi$ is constant during any perturbations to baryon properties.

The stability criterion can then be expressed as
\begin{equation}
    q_{\chi}\!\left(\frac{1}{2m_{\rm B}}{\left(\frac{T_\chi}{m_\chi} + \frac{T_b}{m_{\rm B}} \right)}^{-1} \!\!-\frac{T_\chi}{T_b^2}\left(\frac{T_\chi}{T_b}-1 \right)^{-1} \right)< -\frac{3}{2}q_{\rm rad} T_b^{-1}.
\end{equation}
Since we are performing linear stability analysis around a thermal equilibrium point, $q_\chi = q_{\rm rad}$ and these two factors can be divided out of the inequality.  To simplify the notation, we define a dimensionless baryon temperature $\tau_{\rm B} \equiv T_b/T_{\chi}$ ($\tau < 1$ is necessary for heating) and a dimensionless $\chi$ particle mass $\mu \equiv m_{\rm B}/m_\chi $. The condition for inequality can be rewritten as a simple quadratic expression,
\begin{equation}
    \tau_{\rm B}^2 + \tau_{\rm B} \left(-\frac12+\frac{3}{4}\mu \right) - \frac{\mu}{4} > 0.
\end{equation}
This equation has one unphysical root ($\tau_{\rm B} < 0$) and one physical one, which we label as $\tau_{\rm crit}$.  When $\tau_{\rm B} > \tau_{\rm crit}$, a heating-cooling equilibrium is thermally stable to linear isobaric perturbations; when $\tau_{\rm B} < \tau_{\rm crit}$, it is unstable.  Interestingly, the solution
\begin{equation}
    \tau_{\rm crit} = \frac{2-3\mu +\sqrt{9\mu^2+4(\mu+1)}}{8}
\end{equation}
is monotonic in $\mu$ and has the property of $1/3<\tau_{\rm crit}<1/2$. In the asymptotic limit where $\mu\to 0$, $\tau_{\rm crit} \to 1/3$.
It has been noted previously in the literature (\citet{cavagnolo09}) that the observed baryonic temperature is rarely found to be $\tau_{\rm B} <1/3$. 
Thus the heating is expected to be linearly stable. 
For $n=-2$ and $n=-4$, the condition becomes weaker:
\begin{equation}
    \tau_{\rm crit}=\frac{-3\mu+\sqrt{\mu(9\mu+8)}}{4}
\end{equation}
for $n=-2$ and
\begin{equation}
    \tau_{\rm crit}=\frac{\mu}{3\mu+2}
\end{equation}
for $n=-4$.

\section{Conclusions}
\label{sec:conclusions}

In this paper, we have examined whether novel DM physics has the potential to solve the cooling flow problem in galaxy clusters.  We have built on earlier work by \citet{Qin:2001} and \citet{Chuzhoy:2006} that first proposed baryon-DM heating as a solution to cooling flows, by considering a broader range of DM physics and by comparing heating models to (i) a modern set of astroparticle constraints and (ii) a sample of empirical galaxy cluster profiles.  Our overall approach has been to determine the regions of parameter space that allow DM heating to counterbalance bremmstrahlung cooling in the ICM, ideally in a thermally stable way (as a thermally unstable DM-heated equilibria may develop its own cooling flows).

Our primary conclusions are as follows:
\begin{itemize}


\item DM-baryon scattering models are promising resolutions of the cooling flow problem.  A single-species model of DM-baryon heating is in strong tension with CMB constraints, and we therefore require (at least) a two-component DM model.  In our fiducial model, the large majority of cold DM has no significant interaction with the baryonic sector but serves as a thermal reservoir in galaxy cluster halos.  A second species of DM interacts strongly both with baryons and with the primary DM species, and serves as a mediator to transfer thermal energy between the two. However, there are subtleties related to this fiducial model that require an additional physical assumption (to address CMB constraints), and we propose three such assumptions in Appendix \ref{apx_avoid_cmb}.

\item When the mediator species interacts with baryons through two-body scatterings with a velocity-independent cross-section, a large parameter space exists for both solving the cooling flow problem and evading all existing constraints on baryon-DM interactions.  The two velocity-dependent cross-sections we examined appear to be ruled out by observations of the circum-galactic medium in dwarf galaxies.

\item In contrast with past work, we find that baryon-DM scattering is often a thermally stable heat source, even for energy-independent cross-sections.  Baryon-DM scattering will only become thermally unstable for thermal equilibria (i.e. ICM configurations where heating rates balance cooling rates) with $T_b \ll T_{\chi}$. The narrow and near-unity range of observed $T_b/T_{\rm vir}$ in CC clusters suggests that DM heating may be thermally stable in astrophysical galaxy clusters.

\end{itemize}

Although we have identified a substantial parameter space over which DM-baryon scattering can resolve the cool core problem, there are a number of caveats worth noting.  First, we have only considered the results of DM heating in a relatively simplistic way, by comparing the total energy input to the total bremmstrahlung cooling rate.  A more detailed comparison to data (and test of this model) would be to examine heating models with a coordinate or time dependence, either analytically or in ICM simulations.  Second, a comprehensive survey of the entire landscape of possible DM-baryon interactions is beyond the scope of this work, and it is possible that other heating mechanisms may be more favorable than those studied here.  Third, even if DM-baryon interactions are the dominant ICM heat source in CC clusters, these will coexist with more astrophysical effects, such as AGN heating and the multiphase nature of the ICM.  The interaction between DM-baryon heating and these astrophysical complications has not been considered here, but in principle could lead to useful observational or internal (self-consistency) tests of this model.  These caveats are all areas of investigation we hope to resolve in future work.

\section*{Acknowledgments}
We would like to thank William Forman and Felipe Andrade-Santos for useful discussions and for providing us the {\it Chandra} X-ray profiles used in this work.  We also thank Avishai Dekel and Glennys Farrar for useful discussions and comments on the manuscript. We would like to thank Chingam Fong for pointing out errors in the figure.

Y.S. is supported by the US-Israeli Binational Science Foundation (grant No. 2020220) and the Israel Science Foundation (grant No. 1818/22). E.K. is supported by the US-Israeli Binational Science Foundation (grant 
 2020220), the Israel Science Foundation (grant No. 1111/17).  N.C.S is supported by the Israel Science Foundation (Individual
Research Grant 2565/19) and the Binational Science foundation (grant Nos. 2019772 and 2020397).
Y.B. is supported by the Israel Science Foundation (grant No. 2190/20)

\section*{Data availability}
The data underlying this article will be shared on reasonable request to the corresponding author.

\bibliographystyle{mnras}
\bibliography{cluster_refs,yuval} 

\appendix
\section{CMB Constraints}\label{apx_avoid_cmb}
In order for the DM to heat galaxy clusters, a large coupling between the $\chi$ and the protons is needed. Since $\chi$ behaves like an extra baryonic component, it is severely constrained by CMB measurements. Early universe coupling between the DM and baryonic fluids can leave imprints on the CMB power spectrum and alter baryon acoustic oscillations and the matter power spectrum~\citep{Chen:2002yh}. 

When $f_\chi\lesssim0.4\%$, the density of $\chi$ is within the uncertainty of the baryon density and the constraints {\it can} disappear~\citep{Boddy:2018wzy}. Essentially, this sub-component of DM is cosmologically indistinguishable from the baryons. However, to keep the temperature of $\chi$ above the baryon temperature in cluster cores, the relatively rare $\chi$ particle must be able to access an external heat reservoir; otherwise whatever initial thermal energy is possessed by the $\chi$ would be quickly lost to baryon scattering, and it would not provide a long-term solution for the cooling flow problem.
In our fiducial scenario, where a strong short-range $\psi$-$\chi$ interaction heats the $\chi$ particles, there is a danger of reintroducing the CMB constraint. We found that the high-$z$ energy interchange between $\psi$ and baryons (mediated by $\chi$ particles) is efficient enough to exclude all the parameter space that can solve the CC cluster cooling flow problem. In the following, we provide three additional physical ingredients that can let $\chi$-baryon heating evade the CMB constraint.  While these sub-scenarios should be developed further in future work, each seems to be independently plausible solutions for the CMB constraint.
\subsection{Dark Phase Transition}
The interactions between $\chi$ and $\psi$ at the present time can be different from those at the recombination era. This can be realized, for example, by changing the mediator mass of the interaction through a phase transition in the dark sector \citep{Elor:2021swj}. Then, $\chi$ behaves like baryons and the CMB constraint is evaded if $f_\chi\lesssim0.4\%$.
Conversely, we can consider another case where the interaction between $\chi$ and baryons changes. In this case, $\chi$ behaves like DM and CMB constraints may be avoided even for $f_\chi =100\%$.

\subsection{Particle Decay}
We consider a parent particle, $\chi'$, that decays into $\chi$ and another light particle after recombination. The mass difference between $\chi'$ and $\chi$ is small enough so that the released energy is negligible compared with the baryonic energy at the time of decay.
We assume $\chi'$ has the same charge as $\chi$, but does not have any strong short-range interaction with $\psi$. In addition, we assume $\chi$ does not exist at the recombination era.
Then, $\chi'$ behaves just as baryons until it decays, and the CMB constraint is avoided so long as $\Omega_{\chi'}/\Omega_{\rm DM}\lesssim0.4\%$.
On the other hand, if $\chi'$ does not interact with baryons, then there are no CMB constraint.
\subsection{Dark Conductivity}
Even if $\chi$ and $\psi$ only interact with a weak long-range force and thus cannot exchange energy effectively, $\chi$ particle scatterings with baryons outside the cluster core can still heat $\chi$. In the CC clusters, the baryon temperature in the core is lower than that outside. Thus, $\chi$ transfers energy inwards and heat the core region.
In this sub-scenario, the $\chi$ particles solve the cooling flow problem in much the same way that electron thermal conduction was once thought to \citep{Voigt:2002, Zakamska:2003}; the key difference is that the motion of $\chi$ particles is not inhibited by magnetic fields in the ICM.
Then, the CMB constraint is avoided so long as $f_\chi\lesssim0.4\%$.
In this case, the preferred cross section can be estimated using our formalism for two component DM, but setting $m_\chi\simeq m_p$. Thus the curves would look similar to the left panel in Fig.~\ref{fig:nucleus} but without the sharp rise for low $m_\chi$.

\section{Other External Constraints}
\label{sec:constraints}

\subsection{Collider Constraints}
Collider searches for missing energy can put strong constraints on the production of stable neutral particles that do not interact with the detector. However, for large enough DM-nucleon interactions, the DM will no longer appear as missing energy, and will leave energy in the calorimeter, with a signature similar to a neutron. \cite{Daci:2015hca} find that collider searches for missing energy have a ceiling on the maximum observable DM-nucleon cross-section corresponding to
\beq
\bar{\sigma}_{0}^{\chi N} > 5.5 \times 10^{-27} \frac{\mu_{\chi N}^2}{m_N^2}\,.
\eeq
Cross-sections larger than this value will not be constrained from missing energy searches.

\subsection{Direct Detection}
The majority of DM direction experiments operate underground, while some performed at the Earth's surface, and few are performed in the atmosphere. For very large DM-baryon scattering suggested by the cool-core puzzle, the flux of DM in the experiment is often highly suppressed due to shielding for the atmosphere, Earth's crust, or shielding of the experiment itself. 
The underground direct detection experiments looking for nuclear scattering cannot probe the DM-baryon interactions studied here. The above-ground runs of CREST-III~\citep{CRESST:2017ues} and EDELWIESS~\citep{EDELWEISS:2019vjv} provide the strongest constraints and are sensitive to spin-independent cross-sections up to $\sigma_{\chi N}<10^{-26}~\rm cm^2$. However, these  cross-sections are just below those needed explain the cool-core puzzle. See~\cite{CDEX:2021cll}, for a recent summary of Earth shielding effects on DM direct detection exclusion limits. These constraints are labeled as ``Direct Detection'' in Fig. \ref{fig:nucleus}.

The X-ray Quantum Calorimetry (XQC) experiment \citep{McCammon:2002gb} is a rocket-mounted calorimeter designed to detect x-rays. Although it was not designed for DM detection,   it was an advantage in searching for energy deposits of strongly interacting DM  due to low atmospheric overburden and limited shielding from the rocket. Constraints on the scattering cross section have been derived in the various works~(\citealt{Wandelt:2000ad,Zaharijas:2004jv,Erickcek:2007jv,Mahdawi:2017cxz}). The upper limits on the cross section (lower boundary) are taken by naively rescaling by $f_\chi^{-1}$. The exclusions we use are taken from~\cite{Erickcek:2007jv} whose boundaries are designated by the solid curve. These constraints were more recently updated by~\cite{Mahdawi:2017cxz}, and we show their updated constraints via the dashed curve. Since~\cite{Mahdawi:2017cxz} did not update the ceiling due to shielding, we show both constraints from both works.

We also show constraints from  \cite{Rich:1987st}, a silicon semiconductor detector mounted balloon experiment. These constraints, denoted by RSS, overlap with the XQC experiment.

Additional bounds exist from DM being accelerated by cosmic rays ~\citep{Yin:2018yjn,Bringmann:2018cvk, Ema:2018bih, Cappiello:2019qsw,Ema:2020ulo} or the Sun~\citep{An:2017ojc,Emken:2021lgc}. The boosted DM can then be detected by dark matter and neutrino experiments. For $f_\chi <0.4\%$, the relevant constraints are from XENON-1T~\citep{XENON:2018voc} derived by~\citet{Bringmann:2018cvk}, KamLAND~(\citealt{KamLAND:2011fld}) derived by~\citet{Cappiello:2019qsw}, PandaX-II~(\citealt{PandaX-II:2021kai}) and PROSPECT~(\citealt{PROSPECT:2021awi}). Only PROSPECT is sensitive to cross sections above  $10^{-28}~\rm cm^2$, however for $f_\chi \le 0.05\%$ the experiment loses sensitivity.

\subsection{Cosmic Ray Spectrum}
Strong constraints on the parameter space can be derived from the so called ``reverse direct detection''~\citep{Cappiello:2018hsu}. If DM has strong interactions with baryonic matter, the DM fluid can effect cosmic ray propagation and change the cosmic ray spectrum. These constraints can be powerful for lighter DM, where the abundance is high. 

\subsection{Invisible Meson Decays}
A very strong constraint on the parameter space could arise from the search for invisible meson decays, in particular ${K\to \pi+ { inv}}$.
These constraints are more model-dependent, as they depend on the mediator of the DM-nuclean interaction. The contribution to the kaon invisible width has the possibility to exclude all relevant parameter space, $m_\chi < m_K -m_\pi$. However, for large enough DM-nucleon cross sections, the DM may be visible to the detectors measuring the invisible width of the kaon, such as the NA62 experiment~\cite{NA62:2021zjw}. It may be that no constraint can be obtained from invisible meson decays at these large cross sections. 

\subsection{Leo-T and Milky Way Gas Clouds}
The heating mechanism for the CC clusters can also heat other astronomical objects. We follow \cite{Wadekar:2019mpc} and consider the strongest constraints, which come from Leo-T and Milky Way gas clouds.

Leo-T is a gas-rich dwarf galaxy.
The heating rate of Leo-T is given by Eq.~\eqref{eq_energy_transfer}. We concentrate on $r<1.35~{\rm kpc}$, where the baryon temperature is $6000~\rm K$ and the average radiative cooling rate is $dQ/dt=-3.8\times10^{-30}~{\rm erg~cm^{-3}~s^{-1}}$. The DM velocity is close to that of baryons and
\begin{equation}
    T_\chi-T_b\simeq\frac{(m_\chi-m_p)}{3}\langle v_H^2\rangle.
\end{equation}

For a Milky Way gas cloud, we cannot use Eq.~\eqref{eq_energy_transfer} since the relative velocity between the DM and the gas is not negligible. The energy transfer at the rest frame of the cloud is
\begin{equation}
    \frac{dQ_{\chi\to b}}{dt}=\sum_B\frac{m_\chi\rho_\chi\rho_B}{(m_\chi+m_B)^2}\sigma_n^{\chi B}\Xi_n({\bf V}),
\end{equation}
where $\bf V$ is the relative velocity between the baryons and $\chi$. Here,
\begin{align}
    \Xi_n({\bf V})=A_n\sigma_v\sqrt{\frac{2}{\pi}}e^{-\frac{|{\bf V}|^2}{2\sigma_v^2}}+\frac{B_n}{|{\bf V}|}{\rm Erf}\left(\frac{|{\bf V}|}{\sqrt{2}\sigma_v}\right),
\end{align}
with
\begin{align}
    A_n&=
    \begin{cases}
    |{\bf V}|^2+5\sigma_v^2,&n=0\\
    1,&n=-2\\
    0,&n=-4
    \end{cases},\\
    B_n&=\begin{cases}
    |{\bf V}|^4+6|{\bf V}|^2\sigma_v^2+3\sigma_v^4,&n=0\\
    |{\bf V}|^2+\sigma_v^2,&n=-2\\
    1,&n=-4
    \end{cases}.
\end{align}
The expressions for $n=0,-4$ are available in the Github repository of \cite{Wadekar:2019mpc} and that for $n=-2$ can be obtained by using the formula in the repository. Notice that the expressions in the repository is slightly different from the (D4) of \cite{Wadekar:2019mpc} since they ignore the velocity of baryons.
We consider the G33.4-8.0 cloud, which gives the strongest and most robust constraint as discussed in \cite{Buen-Abad:2021mvc}.
We assume an NFW profile, with  $\rho_{\rm DM}=0.64~{\rm GeV/cm^3}$, $|{\bf V}|=220~{\rm km/s}$, $\sigma_v=124~{\rm km/s}$ and $n_H=0.4~{\rm cm^{-3}}$. We take into account H, He, O, C, Ne and Fe and their mass densities are $\rho_{\rm He}/\rho_{\rm H}=0.33$, $\rho_{\rm O}/\rho_{\rm H}=8.53\times10^{-3}$, $\rho_{\rm C}/\rho_{\rm H}=3.00\times10^{-3}$, $\rho_{\rm Ne}/\rho_{\rm H}=2.30\times10^{-3}$ and $\rho_{\rm Fe}/\rho_{\rm H}=1.62\times10^{-3}$. The radiative cooling rate is $dQ/dt=-1.46\times10^{-27}~{\rm erg~cm^{-3}~s^{-1}}$~\citep{Wadekar:2019mpc}.

\section{DM self interactions and dark charge}
\label{ref:DMself}

Assumed throughout is that the two DM species, $\chi$ and $\psi$, are in kinetic equilibrium and have the same temperature.  This requires a large scattering cross-section between the two species. 
In addition, when $\chi$ is lighter than $\psi$ and they are in the kinetic equilibrium, the velocity dispersion of $\chi$ becomes larger than that of $\psi$. For $\chi$ to be kept in the cluster, the velocity of $\chi$ should not exceed the escape velocity.

One simple model to address this problem is to consider a dark electromagnetic $U(1)$ force where $\chi$ and $\psi$ have opposite sign dark charges. This can introduce a large interaction between the dark particles. Additionally, the cluster is forced to  remain locally neutral because of the long ranged dark Coulomb force, even if the $\chi$ thermal velocity exceeds the escape velocity. 

Without loss of generality, we take the charge of $\psi$ to be $+1\cdot e_D$ and the charge of $\chi$ to be $-q_\chi \cdot e_D$, where $e_D$ is the dark  $U(1)$ gauge coupling.
The relative densities of the two DM particles are fixed by the neutrality of the cluster and we have
\begin{equation}
 f_\chi = \frac{\rho_\chi}{\rho_\psi} = q_\chi^{-1} \frac{m_\chi}{m_\psi}.
\end{equation}

\subsection{Kinetic Equilibrium}
We consider the kinetic equilibrium between $\chi$ and $\psi$ so that they have the same temperature. 
We require the heating of $\chi$ should be faster than that of baryons. The energy transfer is given by Eq.~\eqref{eq_energy_transfer} with replacements; $\chi\to\psi$ and $B\to\chi$.

We assume that kinetic equilibrium is kept by the dark $U(1)$ symmetry and its momentum-transfer cross section is
\begin{equation}
    \sigma_{-4}^{\chi\psi}=\frac{2\pi\alpha_D^2q_\chi^2}{\mu_{\chi \psi}^2}\ln\frac{4\mu_{\chi\psi}^2v_{\rm rel}^2}{\tilde m_D^2},
\end{equation}
similarly as in Eq.~\eqref{sigma-4}.
Here, the dark Debye mass is
\begin{equation}
    \tilde m_D^2=\frac{4\pi \alpha_D}{T_\psi}(n_\psi+q_\chi^2n_\chi).
\end{equation}

Then, requiring that the DM stay in equilibrium imposes the condition
\beq
\frac{\pi q_\chi^2\alpha_D^2/T_\psi^2}{ 2\sigma_0^{\chi N}} \ln\frac{4\mu_{\chi\psi}^2v_{\rm rel}^2}{\tilde m_D^2}\left(\frac{m_B}{m_\psi}\right)^2\left(\frac{ \mu_{\chi\psi}}{ \mu_{\chi N}}\right)^{3/2} \frac{\rho_\psi}{\rho_N} >1.
\eeq
Choosing $\sigma_0^{\chi N}$ to match the baryonic cooling rate we can find an estimate for typical values of $\alpha_D$ required
\beq
\alpha_D \gtrsim 3\times10^{-6}\left(\frac{f_\chi}{0.001\%}\right)^{1/2}\left(\frac{m_\psi}{3~{\rm GeV}}\right)^{2}\left(\frac{0.1~{\rm GeV}}{m_\chi}\right)^{3/4} \label{eq:qalpha}
\eeq
for $m_\chi\ll m_\psi$, $|Q_{\rm rad}|=10^{-31}\rm erg/cm^3/s$, $T_\chi=10^7\rm K$ and $\rho_\psi=10^{-27}\rm gram/cm^3$.

\subsection{Dark Bound States}
For heavier DM and a larger dark charge, the DM may form bound states and the scattering through the dark photon becomes ineffective. We avoid such a region to simplify our discussion.

The binding energy of the DM is
\begin{equation}
    B_\chi\sim\frac{\mu_{\chi\rm DM}q_\chi^2\alpha_D^2}{2},
\end{equation}
where $\alpha_D=g_D^2/(4\pi)$.
To avoid bound states, we require
\begin{equation}
    B_\chi\lesssim T_\chi.
\end{equation}
In addition, we require $\alpha_D q_\chi^2\lesssim1$ to keep the $\chi\chi$ scattering perturbative. We find that this constraint is most significant for $m_\chi \gtrsim$~GeV, where other constraints are already very strong. 

\subsection{Bullet Cluster and Dwarf Evaporation}
A large self coupling of $\psi$ is constrained by the observations of the Bullet Cluster and dwarf galaxies.

From the bullet cluster \citep{Markevitch:2003at,Kahlhoefer:2013dca}, we have
\begin{equation}
    \frac{\alpha_D^2}{m_\psi^3}<550~{\rm GeV^{-3}}.
\end{equation}
From the lack of dwarf evaporation \citep{Kahlhoefer:2013dca}, we have
\begin{equation}
    \frac{\alpha_D^2}{m_\psi^3}<10^{-11}~{\rm GeV^{-3}}.
\end{equation}
Using Eq.~\eqref{eq:qalpha}, this constraint can be quite strong for smaller $m_\chi$ where the parameter space is more open.

These constraints on $\alpha_D$ can be easily addressed if there is an additional contact interaction between $\chi$ and $\psi$. The new interaction keeps kinetic equilibrium within the dark sector, and the dark $U(1)$ interaction keeps $\chi$ in the cluster. Then much smaller $\alpha_D$ than that in Eq.~\eqref{eq:qalpha} is possible, and the constraints discussed in this appendix do not apply.

\section{Light Dark Photon Heating}
\label{sec:dp}
Another attractive heating mechanism appears in light kinetically mixed dark photon DM models. As discussed in \cite{Dubovsky:2015cca}, if a dark photon mixes with the standard model photon, it can create oscillating electric fields within the halo, which accelerate electrons and heat up the plasma. In this mechanism, the energy transfer rate can sometimes be proportional to $n_e^2$ rather than to $n_e$, which is generally favorable for stability to isobaric perturbations (\S \ref{sec_stability}). In the following, we assume that the dark photon constitutes the entire DM population.

We briefly summarize the important formulas below, but refer to \cite{Dubovsky:2015cca,Wadekar:2019mpc} for further details. The relevant Lagrangian is
\begin{align}
    \mathcal L&=-\frac14F_{\mu\nu}F^{\mu\nu}-\frac14\tilde F_{\mu\nu}\tilde F^{\mu\nu}+\frac{m_{\tilde A}^2}{2}\tilde A_\mu \tilde A^\mu\nonumber\\
    &\hspace{3ex}-\frac{e}{(1+\epsilon^2)^{1/2}}J^\mu(A_\mu+\epsilon \tilde A_\mu),
\end{align}
where $A_\mu$ and $F_{\mu\nu}$ are the gauge field and the field strength for the SM photon and $\tilde A_\mu$ and $\tilde F_{\mu\nu}$ are those for the dark photon. The mixing parameter is denoted as $\epsilon$ and the electric current is denoted as $J^\mu$.

The dispersion relation of plasma waves is \citep{Dubovsky:2015cca}
\begin{equation}
    \omega^2=\frac12\left(m_{\tilde A}^2+\Omega_p^2\pm\sqrt{\left(m_{\tilde A}^2+\Omega_p^2\right)^2-\frac{4m_{\tilde A}^2\Omega_p^2}{1+\epsilon^2}}\right).
\end{equation}
Here,
\begin{equation}
    \Omega_p^2=\frac{\omega_p^2}{1+\frac{i\nu}{\omega}},
\end{equation}
with the electron-ion collision frequency
\begin{equation}
    \nu=2\sqrt{2\pi}\frac{\alpha^2n_e}{3m_e^{1/2}T_e^{3/2}}\ln\frac{4\pi T_e^3}{\alpha^3 n_e},
\end{equation}
and $\omega_p =\sqrt{{4\pi n_e\alpha}/{m_e}}$ is the plasma frequency. 

The heating rate is related to the energy loss of the plasma as
\begin{equation}
    \frac{dQ}{dt}=2|{\rm Im}\,\omega|\rho_\chi,
\end{equation}
where
\begin{align}
    {\rm Im}\,\omega&=-\frac{\nu}{2}\frac{\epsilon^2}{1+\epsilon^2}\times
    \begin{cases}
    \frac{m_{\tilde A}^2}{\omega_p^2}&m_{\tilde A}^2\ll \omega_p^2\\
    \frac{\omega_p^2}{m_{\tilde A}^2}&m_{\tilde A}^2\gg \omega_p^2
    \end{cases}.
\end{align}
An isobaric stability analysis (following \S \ref{sec_stability}) shows that dark photon heating is universally stable.  Specifically, the stability criterion $\partial q_{\rm \tilde{A}}/ \partial T_b < \partial q_{\rm rad}/\partial T_b$ becomes  
\begin{equation}
\ln\left(\frac{4\pi T_{\rm e}^3}{\alpha^3n_{\rm e}} \right) > 
\begin{cases}
    0, &m_{\tilde A}^2\ll \omega_p^2 \\
    2, &m_{\tilde A}^2\gg \omega_p^2 \\
\end{cases},
\end{equation}
where we have assumed throughout that $T_b=T_{\rm e}$.  This condition is always satisfied for any astrophysically realistic combination of $T_b$ and $n_{\rm e}$. 

\begin{figure}
    \centering
    \includegraphics[width=\linewidth]{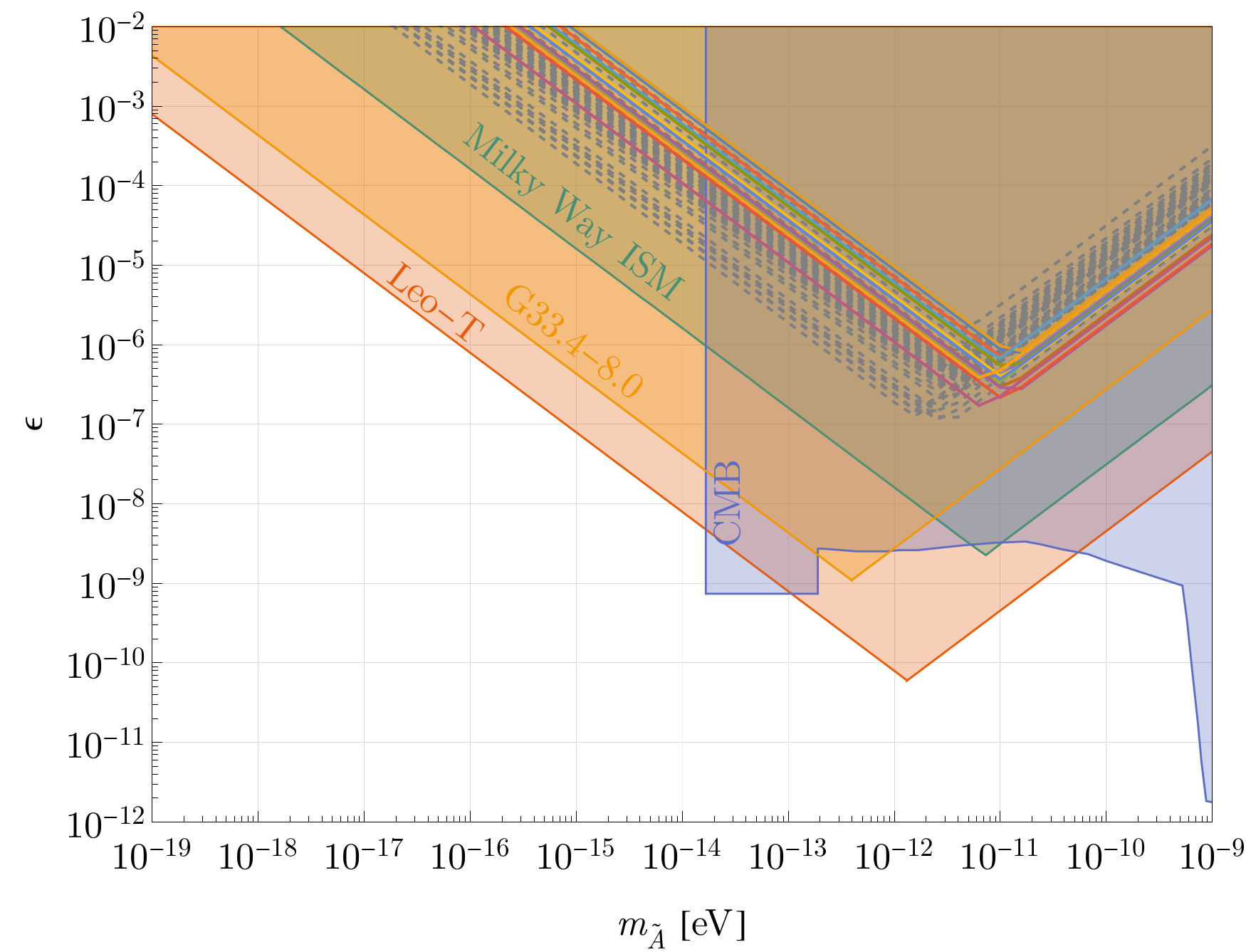}
    \caption{The mixing parameter ($\epsilon$) required for light dark photon DM heating to produce a thermal equilibrium in CC clusters, plotted against dark photon mass $m_{\tilde{A}}$.  Individual curves are as in Fig. \ref{fig:nucleus}.  The equilibria curves for all galaxy clusters overlap multiple exclusion regions from different observational constraints, implying that the dark photon model considered here cannot heat cluster ICM at a high enough rate to prevent cooling flows.}
    \label{fig:darkphoton}
\end{figure}

The parameter space for dark photon heating is constrained by distortions of the CMB (\cite{Mirizzi:2009iz,Arias:2012az}) and heating of the interstellar medium of the MW (\cite{Dubovsky:2015cca}), as well as Leo-T and Milky Way gas clouds (\cite{Wadekar:2019mpc}). Fig.~\ref{fig:darkphoton} shows the constraints on the mixing parameter $\epsilon$ and the dark photon mass $m_{\tilde{A}}$ arising from astrophysical observations, as well as the required $\{\epsilon, m_{\tilde{A}}\}$ for each CC cluster. 
From Fig.~\ref{fig:darkphoton}, we see that any solution to the cooling flow problem from this type of DM-baryon heating is excluded by orders of magnitude by multiple other observations.

\label{lastpage}  
 \end{document}